# Low-Latency Heterogeneous Networks with Millimeter-Wave Communications

Guang Yang, Ming Xiao, *Senior Member, IEEE*, Muhammad Alam, *Senior Member, IEEE*, and Yongming Huang, *Senior Member, IEEE*

*arXiv:1801.09286v1 [cs.NI] 28 Jan 2018*

*Abstract*—Heterogeneous network (HetNet) is a key enabler to largely boost network coverage and capacity in the forthcoming fifth-generation (5G) and beyond. To support the explosively growing mobile data volumes, wireless communications with millimeter-wave (mm-wave) radios have attracted massive attention, which is widely considered as a promising candidate in 5G HetNets. In this article, we give an overview on the end-to-end latency of HetNets with mm-wave communications. In general, it is rather challenging for formulating and optimizing the delay problem with buffers in mm-wave communications, since conventional graph-based network optimization techniques are not applicable when queues are considered. Toward this end, we develop an adaptive low-latency strategy, which uses cooperative networking to reduce the end-to-end latency. Then, we evaluate the performance of the introduced strategy. Results reveal the importance of proper cooperative networking in reducing the end-to-end latency. In addition, we have identified several challenges in future research for low-latency mm-wave HetNets.

## I. INTRODUCTION

### A. Background and Motivation

To significantly improve the spectral efficiency and throughput, future wireless networks, e.g., the fifth-generation (5G) mobile network and beyond, are expected to be largely implemented in the heterogeneous manner, i.e., heterogeneous networks (HetNets) [1]. In HetNets, diverse wireless applications, facility configurations, radio access techniques (RAT), and quality-of-service (QoS) requirements are supported.

With the proliferation of electronic devices and the rapid development of computer science, the traffic load of wireless communications increases continuously and tremendously. To meet the ever-increasing requirements in capacity, one of the most important technologies is millimeter wave (mm-wave), which enables multi-gigabits per second (Gbps) transmission rates, thanks to the abundant spectral resources [2]. Different from conventional mobile communications in sub-6 GHz bands, due to the short wavelength of mm-wave radio, it is easy to integrate tens-to-hundreds of antenna elements onto a small-size chip with lower costs. The resulting high directivity

G. Yang and M. Xiao (corresponding author) are with the Department of Information Science and Engineering (ISE), School of Electrical Engineering (EES), KTH Royal Institute of Technology, Stockholm, Sweden (e-mail: {gy,mingx}@kth.se).

M. Alam is with the Department of Computer Science and Software Engineering, Xi'an Jiaotong-Liverpool University, Suzhou, China (e-mail: alam@ua.pt).

Y. Huang is with the National Mobile Communications Research Laboratory, Southeast University, Nanjing 210096, China (e-mail: huangym@seu.edu.cn).

not only provides a higher antenna gain for combating the severe path loss in mm-wave bands, but also increases the spatial reuse [3]. Besides, due to the fast attenuation of mm-wave signals, the communication distance is commonly limited to short ranges, e.g., 150 to 200 meters. Thus, the inter-cell interference between neighboring small cells is commonly negligible.

In light of above, we notice that mm-wave can be flexibly utilized for diverse devices and network architectures to boost the the coverage and the spectral efficiency. Thus, mm-wave communications has been extensively considered as a promising candidate in HetNets in 5G mobile networks and beyond, and research from various aspects has been massively conducted, e.g., [4], [5]. It is known that, in future mobile communications, latency plays a critical role in the QoS. However, low latency becomes a rather challenging task in 5G mm-wave HetNets, due to the following two facts:

- Buffers will be used in 5G to handle the unprecedentedly heavy traffic, while the queuing delay may seriously deteriorate the QoS in 5G.
- Diverse RATs and/or architectures of HetNets make it rather difficult to perform networking optimizations for lower latency.

Therefore, it is those open challenges that motivate us to investigate the low.latency mm-wave HetNets with buffers in this paper.

### B. Low-Latency Communications

As aforementioned, to support massive and various delay-sensitive applications, low latency as an important QoS feature needs to be satisfied in future wireless communications [6]. In 5G networks, the end-to-end latency requirement will be on the order of 1 to 5 milliseconds (ms) [7], which is more stringent than that in 3G and 4G LTE systems. Thus, it is rather challenging for fulfilling ultra-low latency in future mobile communications.

In the past few years, many efforts have been devoted to low-latency communications. In [8], for ultra-low latency inter-BS communications, the technical challenge and possible solution of point-to-multipoint in-band mm-wave backhaul for 5G networks were studied. In [9], focusing on three critical higher-layer aspects, i.e., core network architecture, protocols at the medium access control (MAC) layer, and congestion control policy, the main challenges and potential solutions for ultra-low latency 5G cellular networks were comprehensively surveyed and discussed. For mm-wave MIMO systems, from the perspective of training time in hybrid beamforming, a novel algorithm based on progressive channel estimation was developed in [10]. In [11], the upper bound on the probabilistic delay was proposed to keep the track of the latency of point-to-point buffer-aided systems with mm-wave.

Considering the unprecedented data volumes in 5G networks, large buffers are usually applied at the transceivers. It is known that, for wireless systems with buffers, the queuing delay dominantly affects the overall system latency [12]. Therefore, for buffer-aided HetNets, it is crucial to realize the ultra-low latency by largely reducing the queuing delay.

### C. Low-Latency HetNets with Buffer

Although many remarkable progresses have been achieved, it is still an open and challenging topic on reducing the end-to-end latency in buffer-aided HetNets. The major difficulty lies in the incorporation of buffers, which makes the problem differ a lot from the conventional latency minimization problems. More exactly, in the presence of buffers, the end-to-end latency relies not only on the capacity of each link, but also on the arriving sequence and queuing state at the buffer. In this sense, the end-to-end latency for buffer-aided networks cannot be simply formulated as a conventional graph-based network optimization problem [13], [14], e.g., shortest path problem, max-flow problem or min-cost flow problem. To the best of our knowledge, the latency minimization problem of the buffer-aided HetNet has not been studied previously.

In this article, we introduce an adaptive strategy for HetNets with buffers, where the cooperative networking is applied. Specifically, we restrict ourselves to a HetNet that consists of one micro cell, two small cells and one user. Results show that the proper cooperative networking plays a critical role in minimizing the end-to-end latency, and our work provides an insight for optimizing future HetNets.

The remainder of this article is organized as follows. In Sec. II, we present the system architecture for 5G HetNets with mm-wave communications, and elaborately discuss several potential scenarios in downlink communications. In Sec. III, we develop an adaptive strategy for minimizing the latency of downlink transmission, based on cooperative networking. In Sec. IV, we evaluate the performance of the introduced adaptive low-latency strategy, which indicates the importance of proper cooperative networking. In Sec. V we identify several technical challenges in future research, and we summarize our work in Sec. VI.

## II. HetNets with mm-wave Communications

A HetNet commonly consists of a macro-cell evolved NodeB (MeNB) and multiple small-cell evolved NodeBs (SeNBs). The MeNB is deployed to guarantee wide-range and seamless coverage, while the SeNBs, e.g., pico, femto, and relay eNBs, are deployed to increase the overall system throughput. In the HetNet with mm-wave communications, as illustrated in Fig. 1, the SeNB is connected to other SeNBs or the MeNB via mm-wave backhaul. The user equipment (UE) gets service from the SeNB via the mm-wave access if it is located in any small cell, and it communicates with the MeNB using microwave radios, otherwise. Thus, for link robustness considerations, dual bands, i.e., mm-wave and microwave bands, are supported at both the MeNB and the UE. It is also possible to have communications working in mm-wave and microwave bands simultaneously, where eleven distinct scenarios need to be considered (with or without the MeNB-UE connection). For analytical simplicity, in this paper, we assume that the UE can only work in either mm-wave bands or microwave bands. In other words, the UE cannot simultaneously connect to the MeNB via the mm-wave link and to the SeNBs via the microwave link.

In what follows, for simplifying illustration, we specifically consider a HetNet that consists of one MeNB, two SeNBs and one UE. In such a network, there are several potential scenarios for downlink transmission from the MeNB to the UE, as illustrated in Fig. 2. These scenarios are elaborated on and discussed as follows:

**Scenario 1:** The UE belongs to neither of the small cells, and it is served by the MeNB via microwave radios, as shown in Fig. 2a. In this scenario, thanks to the direct connection between the MeNB and the UE, there is no extra queuing delay caused by any intermediate node. However, due to the limited bandwidth in microwave bands, the smaller channel capacity (compared to that in mm-wave bands) may produce a larger latency.

**Scenario 2:** As shown in Fig. 2b, the UE communicates with SeNB 2 via the mm-wave access, and SeNB 2 directly connects to the MeNB via the mm-wave backhaul. Thus, a two-hop network is formed. Unlike **Scenario 1**, in spite of two hops, the end-to-end latency can be largely reduced mainly thanks to mm-wave links.

**Scenario 3:** Slightly different from **Scenario 2**, in Fig. 2c, mm-wave backhauls MeNB–SeNB 1 and SeNB 1–SeNB 2 are available, while there is no direct connection between the MeNB and SeNB 2. Thus, the downlink communication is fulfilled through a three-hop network, following the routing MeNB–SeNB 1–SeNB 2–UE.

**Scenario 4:** As shown in Fig. 2d, mm-wave backhaul transmissions are available only for MeNB–SeNB 1 and SeNB 1–SeNB 2. Besides, the UE emerges in the overlapped region of two neighboring small cells, i.e., edge UE, such that it can get served by both SeNB 1 and SeNB 2 via respective mm-wave access, simultaneously. Compared to **Scenario 3**, the difference lies in the UE's association to SeNB 1. In this scenario, the original data traverses from MeNB to SeNB 1. Then, the received traffic at SeNB 1 is divided into two parts: one fraction is directly delivered to the UE from SeNB 1, and the other fraction is delivered to the UE via SeNB 2. Thus, SeNB 2 actually works as a cooperative node, namely, coordinator, which helps SeNB 1 to offload and forward the traffic.

**Scenario 5:** For the edge UE emerging in the overlapped region of two neighboring small cells, if mm-wave backhauls between the MeNB to both SeNBs are available, while SeNBs cannot communicate with each other, then **Scenario 4** becomes **Scenario 5**. In this scenario, the original data traffic is partitioned into two parts at MeNB, which reach the UE via two SeNBs, respectively.

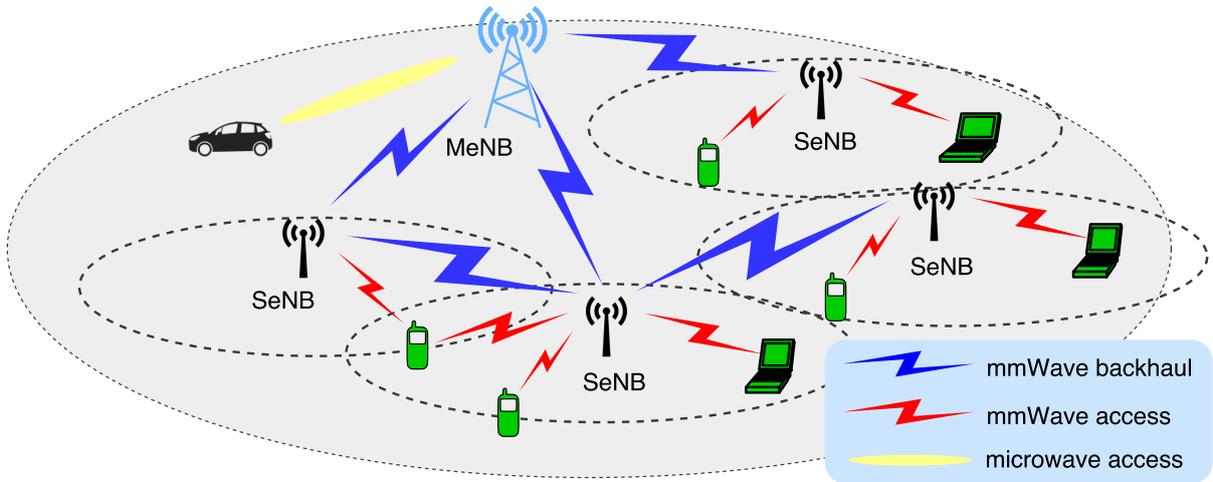

Fig. 1. Illustration of heterogeneous networks (HetNets) with millimeter-wave (mm-wave) and microwave communications.

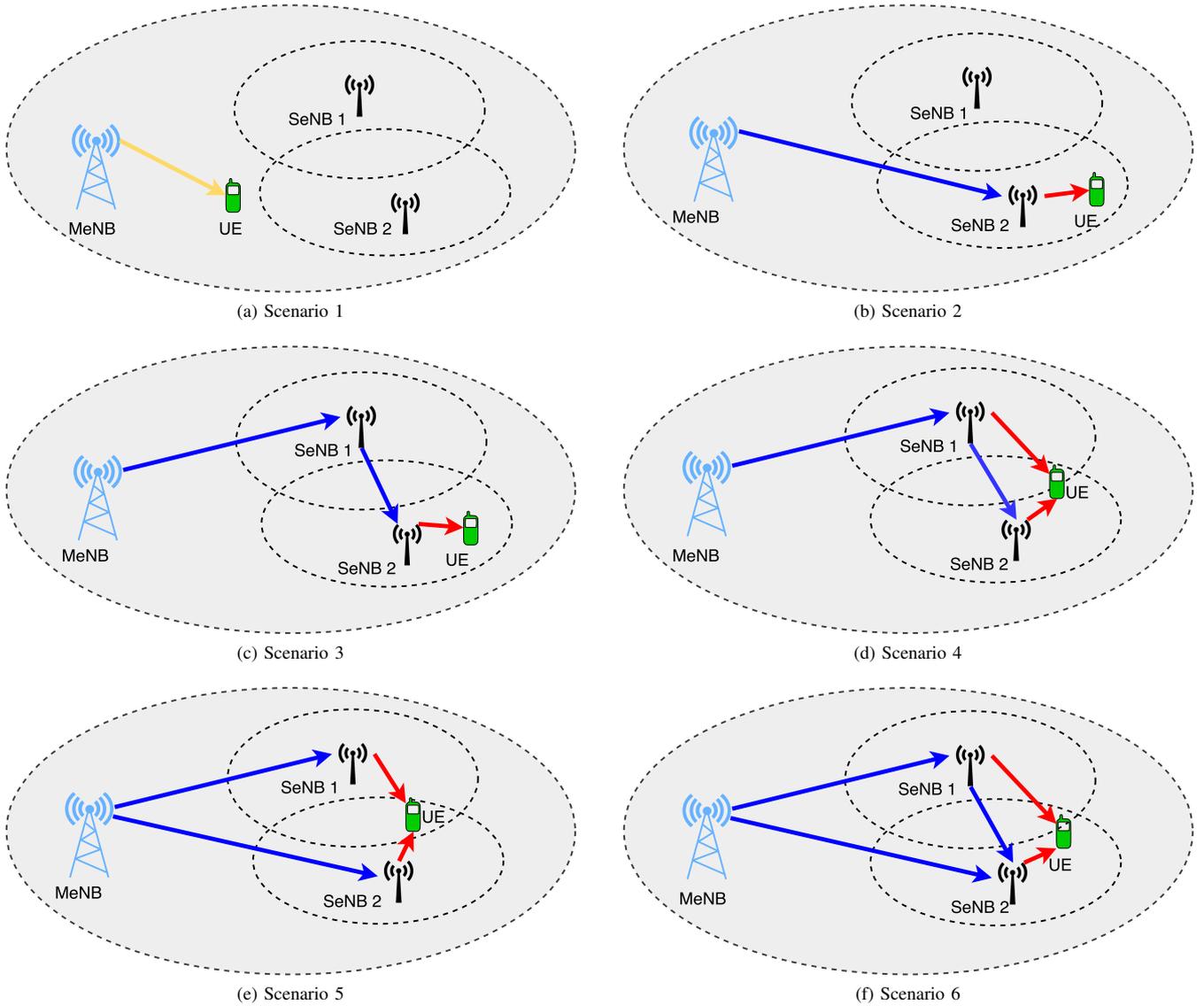

Fig. 2. Potential scenarios for a HetNet with mm-wave communications, where one MeNB, two SeNBs and one UE are considered.

**Scenario 6:** For the edge UE emerging in the overlapped region of two neighboring small cells, if all mm-wave backhauls, i.e., MeNB–SeNB 1, MeNB–SeNB 2, and SeNB 1–SeNB 2, are available, then **Scenario 4** or **Scenario 5** becomes **Scenario 6**. In this scenario, the data traffic may be partitioned and reallocated at the MeNB and SeNB 1, and SeNB 2 as the coordinator is only responsible for merging and forwarding the potential incoming traffic.

For **Scenarios 4** to **Scenario 6**, we notice that there exists the process data splitting and merging. These processes correspond to the *fork-join* system [15] in practice, where the data can be correctly recovered at the UE with synchronization constrains for file transfer.

## III. ADAPTIVE LOW-LATENCY STRATEGY BASED ON COOPERATIVE NETWORKING

### A. Adaptive Low-Latency Strategy

Normally, the MeNB plays the role of "decision maker" or "controller" on the control plane in the HetNet, which determines the networking scheme according to the collected information from SeNBs and the UE. Subsequently, the MeNB, SeNBs and the UE follow the decision distributed from the MeNB, such that the corresponding networking is performed afterwards on the data plane.

We in this section develop an adaptive low-latency transmission strategy in a HetNet with mm-wave communications, which gives the optimal networking scheme according to the acquired channel information. The adaptive strategy is illustrated in Fig. 3, where first-in-first-out (FIFO) queues are used for the buffer-aided HetNet. We can see that there are two planes in the HetNet, namely, the control plane and the data plane. The control plane is responsible for collecting the channel information, directing the networking scheme, and arranging the traffic allocation, where only control signals are operated in this plane. The data plane is only used for data transmission, where all operations are performed under the received control signals. Based on the UE's information, the MeNB will judge if the UE belongs to either small cell. If the UE is outside both small cells, then the MeNB decides to fulfill the downlink transmission in microwave bands. Otherwise, the MeNB needs to design a networking scheme, where the SeNB(s) will potentially participate in the networking. As shown in Fig. 3, data traffic can be partitioned through the traffic splitter, or combined through the traffic merger, under the guidance of the controller. Note that, if the mm-wave backhaul between two SeNBs is available, the controller will consider not only the proper traffic load allocated onto this mm-wave backhaul, but also the proper flow direction, i.e., from SeNB 1 to SeNB 2, or the opposite direction. Thus, both flow directions need to be considered, and the optimal networking is finally made by comparing the potential resulting end-to-end latency.

For $i \in \{1, 2\}$, we denote the channel capacity of mm-wave backhaul between the MeNB and SeNB $i$ by $C_{M,S_i}$,

the channel capacity of mm-wave access between SeNB $i$ and the UE by $C_{S_i,U}$, and the channel capacity of the mm-wave backhaul between two SeNBs by $C_{S_1,S_2}$. Moreover, $\alpha \in \mathcal{A}$ and $\beta \in \mathcal{B}$ represent the traffic allocation coefficients at the MeNB and the non-cooperative SeNB, respectively, where $\mathcal{A} \subset [0, 1]$ and $\mathcal{B} \subset [0, 1]$ are corresponding feasible sets of traffic allocation coefficients. That is, if $\alpha$ (resp. $\beta$) fraction of the file is allocated onto one path, then the left $\bar{\alpha} \triangleq 1 - \alpha$ (resp. $\bar{\beta} \triangleq 1 - \beta$) fraction will be allocated onto the other path. The main idea of the algorithm is that, selecting SeNB 1 and SeNB 2 alternatively as the potential coordinator (the coordinator receives the traffic from both the MeNB and the non-cooperative SeNB, and then forwards the data to the UE), the algorithm traverses each feasible $(\alpha, \beta)$ in the spanned space $\mathcal{A} \times \mathcal{B}$, and computes all potential resulting end-to-end latency. Finally, the proper coordinator, i.e., SeNB $\xi$ with $\xi = 1$ or 2, can be identified, and the optimal allocation pair $(\alpha, \beta)$ can be obtained.

The decision-making procedure at the MeNB is summarized as follows:

1) According to the information of the UE (channel information and location information), the MeNB first judges if Scenario 1 describes the current situation. If yes, a direct transmission in microwave bands will be performed, i.e., $\xi \leftarrow \emptyset$ and $(\alpha, \beta) \leftarrow (\emptyset, \emptyset)$. Otherwise, the downlink transmission requires the participation of SeNB(s), and the MeNB performs the following steps to make the networking decision.

2) Treating SeNB 1 as the coordinator, the MeNB computes the minimum end-to-end latency $\tau_1^*$ and the associated optimal allocations $(\alpha_1^*, \beta_1^*)$, in the presence of known $C_{M,S_1}$, $C_{M,S_2}$, $C_{S_1,S_2}$, $C_{S_1,U}$, $C_{S_2,U}$, $\mathcal{A}$ and $\mathcal{B}$. Meanwhile, treating SeNB 2 as the coordinator, the MeNB computes the minimum end-to-end latency $\tau_2^*$ and the associated optimal allocations $(\alpha_2^*, \beta_2^*)$, likewise.

3) By comparing $\tau_1^*$ and $\tau_2^*$, the MeNB finally selects SeNB $\xi$ with $\xi \leftarrow \arg_{i \in \{1,2\}} \min \tau_i^*$ as the coordinating SeNB, and the corresponding $(\alpha^*, \beta^*) \leftarrow \left(\alpha_\xi^*, \beta_\xi^*\right)$ will be adopted for networking as the optimal traffic allocations.

Note that the end-to-end delay only depends on the routing decision and the link capacities. Thus, the strategy proposed above can work for both multi-tier and multi-RAT HetNets.

### B. Traffic Allocation for Cooperative Networking

The strategy of traffic allocation for cooperative networking is shown in Fig. 4. $\alpha$ and $\beta$ denote the fractions of traffic allocated onto MeNB–SeNB 2 and SeNB 1–SeNB 2 mm-wave backhauls, respectively (SeNB 2 is taken as the coordinator in Fig. 4). Let the size of file for the downlink transmission be $L$ units. As shown in Fig. 4a, the first allocation happens at the MeNB, where $\alpha L$ and $\bar{\alpha} L$ units of the file are pushed onto the mm-wave backhauls MeNB–SeNB 2 and MeNB–SeNB 1, respectively. The second allocation happens at SeNB 1, where the received $\bar{\alpha} L$ units are divided into two parts, i.e., $\bar{\alpha}\beta L$ units and $\bar{\alpha}\bar{\beta} L$ units, for transmissions on the mm-wave backhaul SeNB 1–SeNB 2 and the mm-wave access SeNB 1–UE, respectively. SeNB 2 receives $\alpha L$ and $\bar{\alpha}\bar{\beta}L$ units from both MeNB and SeNB 1, and buffers them in the queue. The downlink transmission is not completed until all units reach the UE, i.e., $\bar{\alpha}\bar{\beta}L$, $\bar{\alpha}\beta L$, and $\alpha L$. An abstraction for this procedure is illustrated in Fig. 4b, where $w_1$, $w_2$ and $w_3$ denote component delays on three traversing paths, respectively, and the largest one among $w_1$, $w_2$ and $w_3$ defines the end-to-end latency.

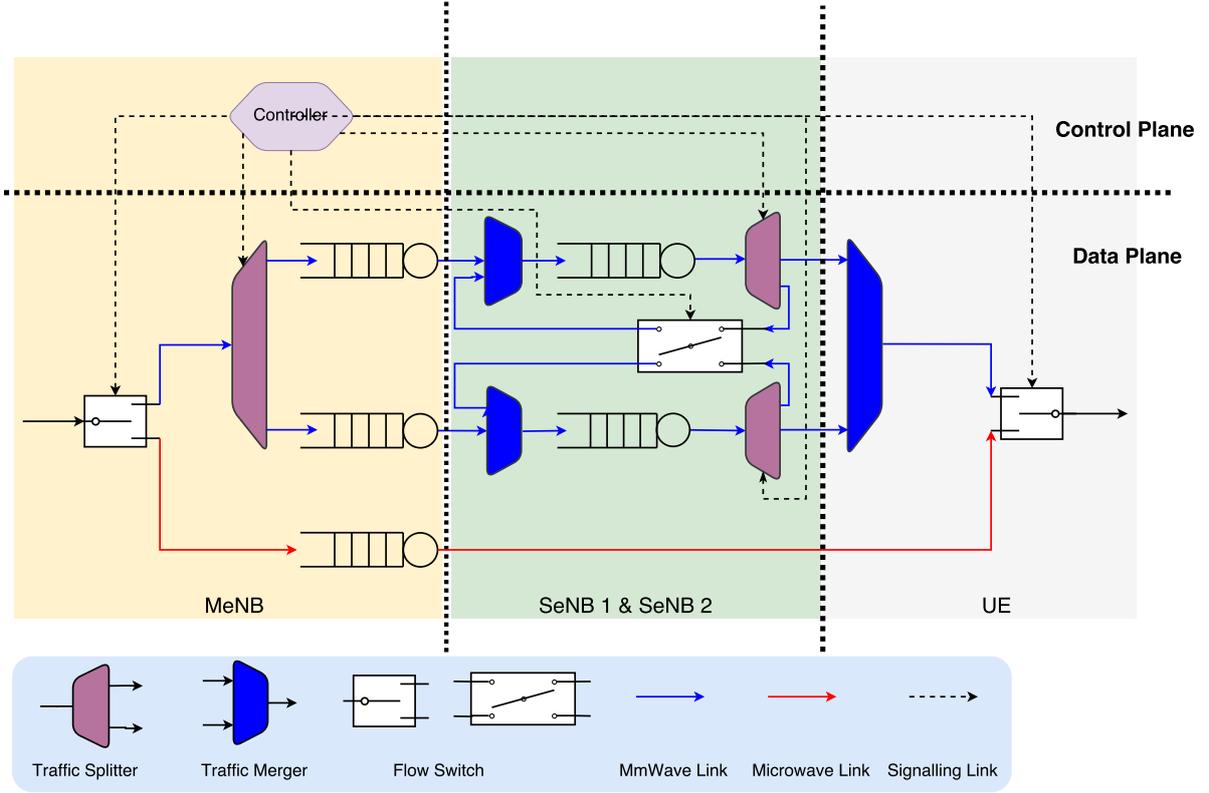

Fig. 3. Diagram of adaptive low-latency strategy, where first-in-first-out (FIFO) queues are used for the buffer-aided HetNet.

We assume that the feasible sets of traffic allocations are given as $\mathcal{A} = \{\alpha : 0 \le \alpha \le 1\}$ and $\mathcal{B} = \{\beta : 0 \le \beta \le 1\}$, respectively. Taking the potential scenarios listed in Fig. 2 for example, the values for $\alpha$ and $\beta$ are correspondingly given as follows:

- $\alpha \leftarrow \emptyset$ and $\beta \leftarrow \emptyset$ for **Scenario 1**
- $\alpha \leftarrow 1$ and $\beta \leftarrow \emptyset$ for **Scenario 2**
- $\alpha \leftarrow 0$ and $\beta \leftarrow 0$ for **Scenario 3**
- $\alpha \leftarrow 0$ and $\beta \leftarrow \beta^* \in (0,1)$ for **Scenario 4**
- $\alpha \leftarrow \alpha^* \in (0,1)$ and $\beta \leftarrow 0$ for **Scenario 5**
- $\alpha \leftarrow \alpha^* \in (0,1)$ and $\beta \leftarrow \beta^* \in (0,1)$ for **Scenario 6**

For minimizing the end-to-end latency, it is crucial to optimize $\alpha$ and $\beta$. Recalling the abstraction in Fig. 4b, for the first routing, i.e., MeNB–SeNB 1–UE, the component delay $w_1$ can be easily formulated. However, for the second routing and the third routing, i.e., MeNB–SeNB 1–SeNB 2–UE and MeNB–SeNB 2–UE, it is necessary to consider the order of arrivals of two distinct file fractions at SeNB 2. To be more precise, with FIFO queuing, the earlier arrival will be pushed onto mm-wave access SeNB 2–UE first, and the later one may have to wait in the queue until the earlier comer completely departs from the buffer. Comparing the arriving order of different file fractions, we then are able to formulate component delays $w_2$ and $w_3$. Finally, to achieve $\tau_1^* \leftarrow \min \max \{w_1, w_2, w_3\}$, the MeNB traverses all feasible $(\alpha, \beta) \in \mathcal{A} \times \mathcal{B}$ to identify the optimal traffic allocations, i.e., $(\alpha_1^*, \beta_1^*)$, which enables the minimal end-to-end latency when treating SeNB 2 as the coordinator.

It is worth mentioning that, the adaptive low-latency strategy based on cooperative networking can be extended to general scenarios with more than two SeNBs, as long as the channel information of all potential links is available for performing the global optimization, since a high-dimensional traffic allocation vector can always be generated for optimization with the global information. However, the major challenge with expanding size of HetNets is the computational complexity at the MeNB, since the cost for processing the channel information of all links and performing global optimization over a full-connected network (counting all potential links) dramatically increases. In this sense, it is necessary to consider the trade-off between the achieved latency and the computational complexity in practice. This issue is identified as a future challenge, stated in Sec. V.

## IV. PERFORMANCE EVALUATION

In this section, we evaluate the performance of cooperative networking for HetNets with mm-wave communications. To

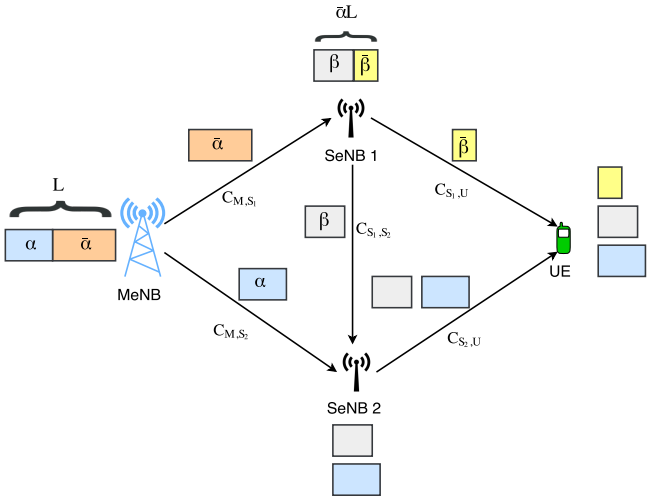

(a) Illustration of traffic allocation and networking procedure for the downlink transmission (from MeNB to UE, via SeNB(s) potentially).

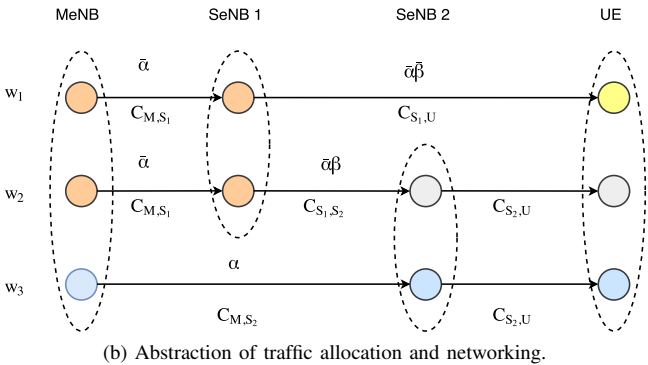

(b) Abstraction of traffic allocation and networking.

Fig. 4. Traffic allocation and networking procedure, and the corresponding abstraction.

investigate the impacts of traffic allocation pair $(\alpha, \beta)$, we assume deterministic settings for HetNets with mm-wave communications. That is, the channel capacities of mm-wave backhauls and accesses are $C_{M,S_1} = 12$ Gbps, $C_{M,S_2} = 8$ Gbps, $C_{S_1,S_2} = 7$ Gbps, $C_{S_1,U} = 0.8$ Gbps, and $C_{S_2,U} = 2$ Gbps, and the size of file for the downlink transmission is $L = 2$ Mb.

With different traffic allocation $(\alpha, \beta)$, the end-to-end latency is shown in Fig. 5, where SeNB 1 and SeNB 2 are selected as the coordinator in Fig. 5a and Fig. 5b, respectively. The region in dark blue represents the desired sets of feasible $(\alpha, \beta)$, which can provide a lower end-to-end latency for downlink transmission. We can see that, in both sub-figures, the dark blue regions emerge inside the square $[0,1] \times [0,1]$ for all potential $\alpha$ and $\beta$, i.e., $\alpha \in \mathcal{A} \setminus \{0,1\}$ and $\beta \in \mathcal{B} \setminus \{0,1\}$. This is resulted by the availability of mm-wave backhaul SeNB 1–SeNB 2, i.e., $C_{S_1,S_2} = 7 > 0$ Gbps. This observation indicates that, if the mm-wave backhaul between SeNBs is available, it is always beneficial to take advantages of this backhaul by properly performing traffic allocations, and the resulting end-to-end latency can be much less than those without traffic allocations, i.e., strategies with $\alpha \in \{0,1\}$ or $\beta \in \{0,1\}$. Furthermore, comparing Fig. 5a and Fig. 5b, we find that the minimum end-to-end latency is $0.396$ ms when SeNB 1 is treated as the coordinator, while the minimum end-to-end latency is $0.426$ ms when SeNB 2 is treated as the coordinator. Therefore, it is a better choice to take SeNB 1 as the coordinator, and perform the optimal traffic allocations at the MeNB and SeNB 2, respectively. The decision above is finally made and will be distributed from the MeNB for the subsequent low-latency cooperative networking.

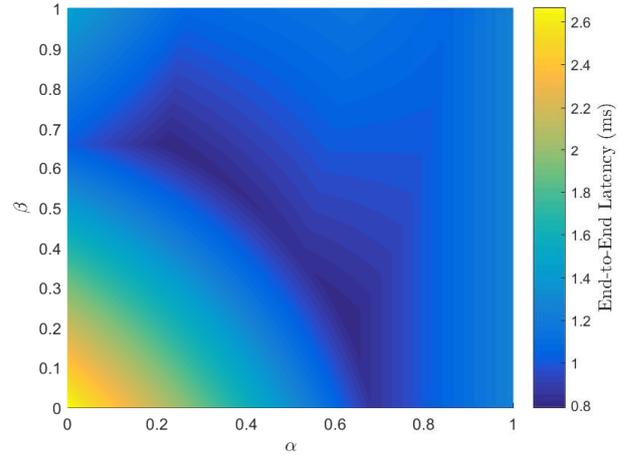

(a) SeNB 1 as the coordinator

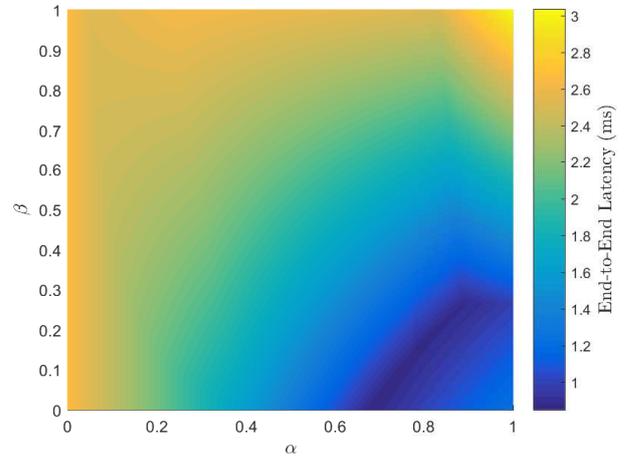

(b) SeNB 2 as the coordinator

Fig. 5. End-to-end latency with traffic allocation coefficient pairs $(\alpha, \beta)$: (a) taking SeNB 1 as the coordinator; (b) taking SeNB 2 as the coordinator.

## V. CHALLENGES IN FUTURE RESEARCH

Due to the densification tendency for small cells in future HetNets, the adaptive low-latency strategy developed in this article may face a few new technical challenges as follows:

- We only consider two small cells in our research. However, the overhead for collecting channel information and control signaling becomes tremendously heavy when more small cells are incorporated.
- The performance of the proposed approach depends on the channel state information. However, at mm-wave frequencies, it might be more difficult to obtain the channel state information due to the severe Doppler effect when mobility is involved. Thus, for mobile scenarios, it is a challenging task to overcome the degradation cased by Doppler effect.

- Since the capacities of mm-wave backhauls or accesses is not infinite, it is critical to properly schedule transmissions and manage the traffic in the presence of multiple UEs in the HetNet, which however is a non-trivial optimization problem.

Thus, our future work will focus on developing a low-complexity and scalable algorithm for low-latency wireless communications in HetNets with buffers.

## VI. Conclusions

HetNets with mm-wave communications can significantly improve the network coverage and capacity, to satisfy ever-increasing requirements in data rates and latency. We have considered a HetNet consisting of one MeNB, two SeNBs and one UE, and investigated the low-latency strategy for the downlink transmission from the MeNB to the UE. For the HetNets with buffers, we have introduced an adaptive strategy based on cooperative networking, which largely minimizes the latency through optimizing traffic allocations. Results have demonstrated that, a proper cooperative networking is critical in reducing the end-to-end latency, thereby providing an insight on traffic management and network optimization for future HetNets. Besides, we have identified several challenges regarding cooperative communications in low-latency HetNets to be addressed in future research.

## Acknowledgment

This work was supported by EU Marie Curie Project, QUICK, No. 612652, and Wireless@KTH Seed Project "Millimeter-Wave for Ultra-Reliable Low-Latency Communications".

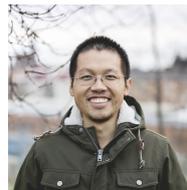

**Guang Yang** received his B.E degree in Communication Engineering from University of Electronic Science and Technology of China (UESTC), Chengdu, China in 2010, and from 2010 to 2012 he participated in the joint Master-PhD program in National Key Laboratory of Science and Technology on Communications at UESTC. He joined the Information Science and Engineering, Electrical Engineering School, Royal Institute of Technology (KTH), Stockholm, Sweden, as a Ph.D. student since September of 2013.

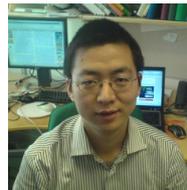

**Ming Xiao** (S'2002-M'2007-SM'2012) received his Ph.D. degree from Chalmers University of Technology, Sweden, in November 2007. From November 2007 to now, he has been in Information Science and Engineering, School of Electrical Engineering, Royal Institute of Technology, Sweden, where he is currently an associate professor in communications theory. He received the best paper awards at the International Conference on Wireless Communications and Signal Processing in 2010 and the International Conference on Computer Communication Networks in 2011.

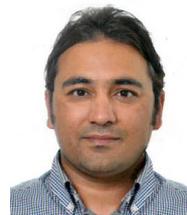

**Muhammad Alam** holds a PhD degree in computer science from University of Aveiro, Portugal (2013-14) and has a MS degree in Computer Science from International Islamic university Islamabad, Pakistan. He has participated in several EU funded projects such as C2POWER, ICSI, Hurricane, and PEACE. Currently, he is working as Assistant professor in Xi'an Jiaotong-Liverpool University, Suzhou China. His research interests include IoT, Real-time wireless communication, 5G and Vehicular networks. He is senior member IEEE.

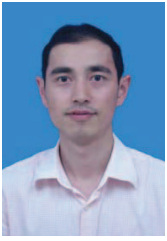**Yongming Huang** (M'10-SM'17) received the B.S. and M.S. degrees from Nanjing University, China, in 2000 and 2003, respectively, and received the Ph.D. degree in electrical engineering from Southeast University, China, in 2007. He is currently a full professor in the School of Information Science and Engineering, Southeast University, China . His current research interests include MIMO communications and millimeter wave communications.